\documentclass[%
 prb,
 twocolumn,
 superscriptaddress,
 amsmath,amssymb,
 aps,
]{revtex4-2}

\usepackage{graphicx}
\usepackage{dcolumn}
\usepackage{bm}
\usepackage{float}
\usepackage{hyperref}
\usepackage{times}
\hypersetup{colorlinks=true, citecolor=blue, urlcolor=blue, linkcolor=blue}

\begin{document}

\preprint{APS/123-QED}

\title{Active orbital degree of freedom and potential spin-orbit-entangled moments in Kitaev magnet candidate BaCo$_2$(AsO$_4$)$_2$}

\author{Subhasis Samanta}
\thanks{These two authors contributed equally.}
 \affiliation{Department of Physics and Institute of Quantum Convergence Technology, Kangwon National University, Chuncheon 24341, Korea}
 
\author{Panyalak Detrattanawichai}
\thanks{These two authors contributed equally.}
\affiliation{
 Division of Physics, School of Science, Walailak University, Nakhon Si Thammarat, 80160, Thailand
}

\author{Sutassana Na-Phattalung}%
\email{sutassana.na@mail.wu.ac.th}
\affiliation{
 Division of Physics, School of Science, Walailak University, Nakhon Si Thammarat, 80160, Thailand
}
\affiliation{
Functional Materials and Nanotechnology Center of Excellence, Walailak University, Nakhon Si Thammarat, 80160, Thailand
}

\author{Heung-Sik Kim}
\email{heungsikim@kangwon.ac.kr}
\affiliation{Department of Physics and Institute of Quantum Convergence Technology, Kangwon National University, Chuncheon 24341, Korea}

\begin{abstract}
Candidate materials for Kitaev spin liquid phase have been intensively studied recently because of their potential applications in fault-tolerant quantum computing. Although most of the studies on Kitaev spin liquid have been done in 4$d$ and 5$d$ based transition metal compounds, recently there has been a growing research interest in Co-based quasi-two-dimensional honeycomb magnets, such as BaCo$_2$(AsO$_4$)$_2$ because of formation of spin-orbit-entangled $J_{\rm eff}$ = 1/2 pseudospin moments at Co$^{2+}$ sites and potential realizations of Kitaev-like magnetism therein. Here, we obtain high-accuracy crystal and electronic structure of BaCo$_2$(AsO$_4$)$_2$ by employing a combined density functional and dynamical mean-field theory calculations, which correctly capture the Mott-insulating nature of the target system. We show that Co$^{2+}$ ions form a high spin configuration, $S=3/2$, with an active $L_{\rm eff}=1$ orbital degree of freedom, in the absence of spin-orbit coupling. The size of trigonal distortion within CoO$_6$ octahedra is found to be not strong enough to completely quench the orbital degree of freedom, so that the presence of spin-orbit coupling can give rise to the formation of spin-orbit-entangled moments and the Kitaev exchange interaction. Our finding supports recent studies on potential Kitaev magnetism in this compound and other Co-based layered honeycomb systems. 
\end{abstract}

\maketitle

\section{Introduction}

Quantum spin liquids (QSLs) are states of matter with disordered yet highly entangled spin moments caused by magnetic frustrations \cite{Ng2017,Savary2016,Nagler2019}. In addition to conventional geometrically frustrated QSLs in triangular or kagome lattices, a new type of spin liquid known as Kitaev QSL has been proposed, where bond-dependent Ising-type anisotropic exchange interactions prevent any long-range magnetic order and induce fractionalized Majorana excitations \cite{Kitaev2006}. Soon after it has been found that such Kitaev magnetism can be realized in crystalline solids with strong spin-orbit coupling \cite{Khaliullin2009,Khaliullin2010}, after which a number of suggestions for candidate materials has followed \cite{Gegenwart2012,Kim2013,Takagi2015,Analytis2014,Kee2014,Kee2015}. Initially, 4$d$-transition-metal-based $\alpha$-RuCl$_3$ and 5$d$-based iridates have been studied as prototypes to realize such novel magnetic states, because spin-orbit coupling, strong correlation, and crystal field-effect altogether give rise to a $J_{\rm eff}=1/2$ state that forms bond-directional exchange interactions in these systems \cite{Kee2014}. However, subsequent studies have shown that there are substantial amount of non-Kitaev interactions present in these materials \cite{Lee2014}. For instance, in $\alpha$-RuCl$_3$ and iridate compounds, the presence of non-cubic octahedral distortions and spatially extended 4$d$- and 5$d$-orbitals result in appreciable non-Kitaev interactions, which break QSL and lead to long-range magnetic orders \cite{Kee2016,Valenti2016}.

Most of the proposed Kitaev materials, studied so far in the literature, largely deviate from the original Kitaev model because of the non-negligible presence of nearest and further-neighbor Heisenberg interactions, symmetric anisotropy exchanges, and contamination by other non-spin-orbit-entangled states \cite{He2021,Nagler2016,Aruga2020}. Because of this, over the past few years there has been a growing interest in 3$d$-based honeycomb materials as potential candidates for the Kitaev magnetism \cite{Khaliullin2020,Lui2021,ParK2020,Park2021}. It has been reported that materials with $d^7$ configurations in the cubic environment can host  spin-orbit-entangled $J_{\rm eff}$ = 1/2 pseudospins \cite{Motome2018,Khaliullin2018}, with less amount of long-range non-Kitaev magnetic interactions compared to their 4$d$ and 5$d$ counterparts.

Among 3$d$-based honeycomb magnets, BaCo$_2$(AsO$_4$)$_2$ (BCAO) has attracted a great deal of research interest recently. A couple of recent experimental studies have proposed BCAO as a candidate closest to the ideal Kitaev spin liquid \cite{Cava2020,Armitage2022}. It has been reported earlier in the literature that there is no magnetic order in BCAO down to 5.4 K. Below critical temperature, it shows a strong field and temperature-dependent magnetization with two successive magnetic transitions at the applied in-plane magnetic field of 0.26 T and 0.52 T \cite{Cava2020}. A single crystal x-ray diffraction study reported that there is no stacking fault or formation of domains in BCAO \cite{Cava2020}. A subsequent time domain spectroscopic study has suggested that BCAO has a dominant Kitaev interaction, where applying small in-plane magnetic fields $\sim$ 0.5 T suppresses the magnetic order \cite{Armitage2022}. Nevertheless, there is an ongoing debate on the nature of the magnetic moments and magnetic exchange interactions and formation of spin liquid phases in BCAO \cite{Paramekanti2021,Streltsov2022,Winter2022,Armitage2022,Cava2020,Wang2021}, yet there has not been a direct ab-intio-based confirmation of the nature of the spin-orbit-entangled moments.

Unlike the formation of the spin-orbit-entangled $J_{\rm eff}=1/2$ local pseudospin moments in compounds like iridates  \cite{Rotenberg2008,Arima2009,Yu2013,Noh2013,Yong2015,Yong2016} or Ru-based trihalides \cite{Kee2016,Andrei2016,Heung2021}, the nature of the spin-orbit-entangled moments in Co-based Kitaev candidate compounds makes them hard to be captured with simple density functional theory calculations because of their inherent multiplet nature of Co $d^7$ configurations \cite{Khaliullin2020,Lui2021,ParK2020,Park2021,Motome2018,Khaliullin2018,Park2022}. To tackle this issue, by employing first-principles density functional and dynamical mean-field theory calculations, we investigate the crystal and electronic structures of BCAO, especially focusing on elucidating the formation of high spin state and orbital degrees of freedom within Co$^{2+}$ ions in BCAO. Our study unveils that despite the presence of sizeable amount of trigonal distortion in CoO$_6$ octahedra, orbital angular momentum is not fully quenched and an active $L_{\rm eff}=1$ orbital degree of freedom survives. Therefore, with the incorporation of spin-orbit coupling in the calculations, the system strongly favors a formation of spin-orbit-entangled $J_{\rm eff}=1/2$ state. The presence of edge-sharing octahedra and anisotropic exchange due to spin-orbit entangled moment, in addition to spatially localized Co $d$-orbitals that suppress longer-range Heisenberg interactions which break spin-liquid phases \cite{Valenti2016}, idealizes BCAO for the realization of Kitaev spin liquid phase.

\section{Computational details}

Density functional theory (DFT) calculations for initial electronic structure calculations and structural relaxations were performed employing a projector-augmented-wave (PAW) basis as implemented in a pseudopotential-based Vienna {\it ab-initio} Simulation Package ({\sc vasp}) code \cite{Kresse1996}. A revised Perdew-Burke-Ernzerhof generalized gradient approximation exchange-correlation functional for crystalline solids (PBEsol) was employed to optimize the crystal structure \cite{Burke2008}. Note that aside from the exchange-correlation functional, we do not include additional on-site Coulomb repulsion effects in DFT calculations, like DFT+$U$ terms. The $k$-space was sampled with a $\Gamma$-centered 7$\times$7$\times$7 mesh. Energy cutoff was set to be 500 eV and a force criterion of 10$^{-4}$ eV/\AA~was employed for internal atomic coordinate optimizations. Note that lattice parameters were adopted from {\sc materials project} database (\href{https://materialsproject.org}{https://materialsproject.org}) (ID: mp-19143) and left unchanged in this study.

Dynamical mean-field theory (DMFT) calculations were carried out using the Embedded DMFT Functional code \cite{Haule2010,Haule2018} interfaced with the full-potential-based {\sc wien2k} code \cite{Blaha2018}. The Cerpeley-Alder parametrization of the local density approximation (LDA) functional was used in the DMFT calculations \cite{Alder1980}. A 7$\times$7$\times$7 $k$-mesh was used for sampling the $k$-space. The results were additionally tested with employing 12$\times$12$\times$12 and 14$\times$14$\times$14 $k$-grids (see App. \ref{ref:App4}). The $RK_{\rm max}$ was taken to be 7. A simplified Ising-type density-density Coulomb interaction was employed, where the results were verified with using full Coulomb interactions (see App. \ref{ref:App2}). The Anderson impurity problem was solved using hybridization-expansion continuous-time quantum Monte Carlo method (CT-HYB) \cite{Warner2006,Haule2007}, where the temperature of the electron bath and the number of Monte Carlo steps for each DMFT iteration were set to be 232 K and 1.6$\times$10$^9$, respectively (we also checked that our main result remains valid at a lower temperature of $T$ = 50 K, please refer to App. \ref{ref:App5} for more details). We chose the full Co $d$ orbitals as the correlated subspace, where two sets of Coulomb parameters, $(U, J_{\rm H}$) = $(8, 0.8)$ and $(12, 1.2)$ eV, were considered. Additionally, potential double-counting dependency of our DMFT result was checked by choosing different choices of nominal $d$-occupancy in the nominal double counting scheme (see App. \ref{ref:App3}). Finally, internal ionic coordinates were fully optimized in DMFT calculations with a force criterion of 12 meV/{\AA} and were compared with PBEsol-optimized ones.

\section{Weakly correlated electronic structure of BCAO from nonmagnetic DFT}

\begin{figure}
\centering
\includegraphics[width=0.5\textwidth]{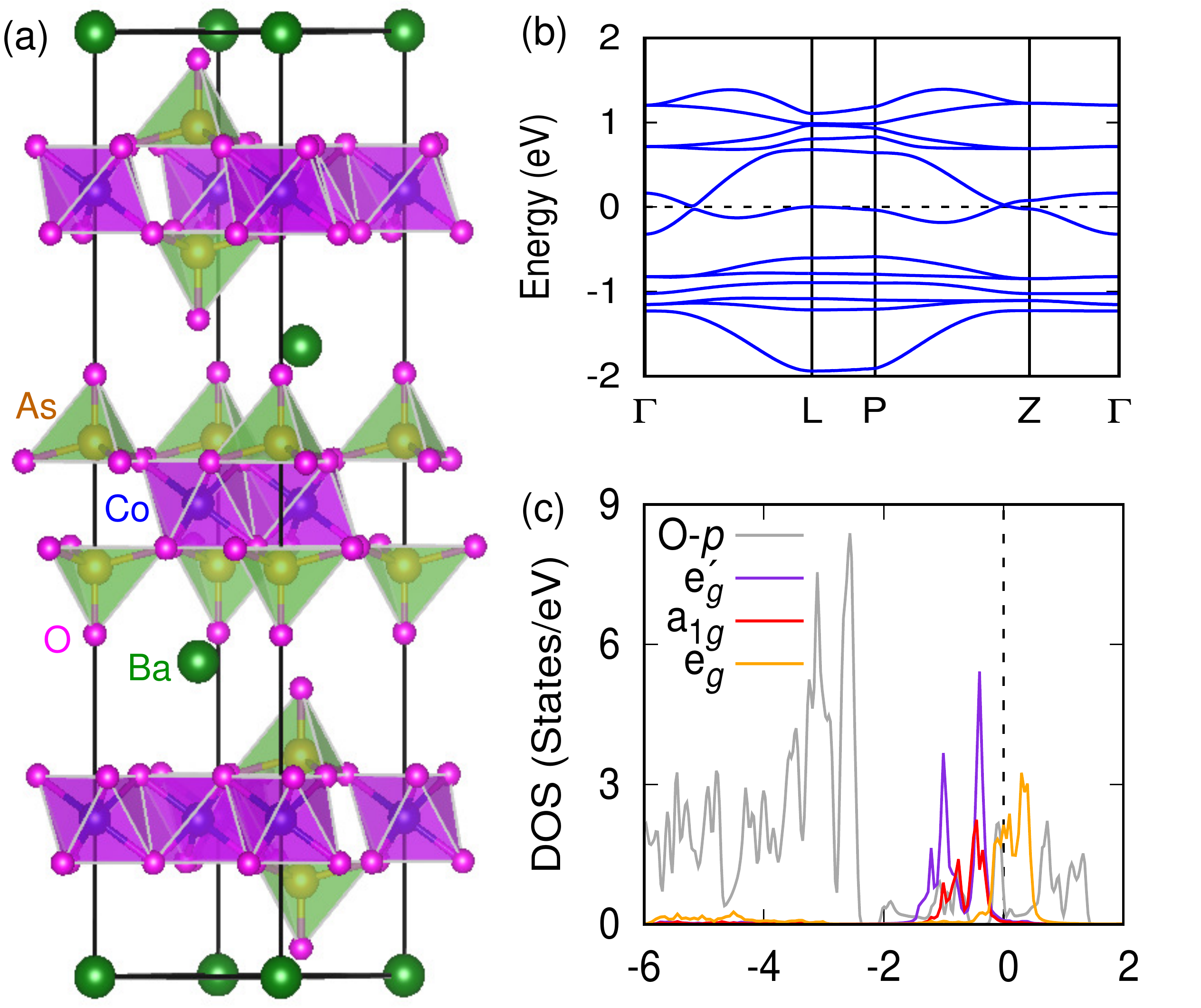}
\caption{(a) Crystal structure of two-dimensional honeycomb magnet BCAO. (b) Nonmagnetic band structure in the primitive Brillouin zone obtained using LDA. (c) Projected density of states of O-$p$ (gray) and Co-$d$ orbitals split into $e^\prime_{\rm g}$ (violet), $a_{\rm 1g}$ (red), and $e_{\rm g}$ (orange).}
\label{fig:Fig1}
\end{figure}

BCAO crystallizes in centrosymmetric rhombohedral structure with a space group $R\bar{3}$. Figure~\ref{fig:Fig1}(a) shows a side view of the crystal structure of BCAO, where Ba cations and AsO$_4$ tetrahedra that cap both sides of CoO$_3$ layers greatly enhance the inter-layer distance and two-dimensionality in this compound. Note that Co honeycomb layers form a rhombohedral stacking pattern. Two adjacent Co honeycomb layers are roughly separated by 8 {\AA}. 
 
In this work, we focus on the nature of the paramagnetic phase of BCAO above $T_c$ = 5.4 K \cite{Mignod1977}. First, we present non-magnetic and non-spin-polarized electronic structure of BCAO obtained from DFT calculations without incorporating spin-orbit coupling, as depicted in Fig.~\ref{fig:Fig1}(b), where the band structure of BCAO along the high symmetry paths in the primitive Brillouin zone is shown. In the presence of trigonal crystal field, Co-$d$ orbitals are further split into $e^\prime_{\rm g}$, $a_{\rm 1g}$, and $e_{\rm g}$ orbitals. The lowermost six bands are derived from Co-$(e^\prime_{\rm g}, a_{\rm 1g})$ orbitals and have nearly flat out-of-plane dispersion along the $\Gamma-Z$ line, implying the two-dimensional nature of BCAO. The bands close to the Fermi level ($E_{\rm F}$) consist of Co-$e_{\rm g}$ orbitals and show noticeable dispersion along $\Gamma$ to $Z$ due to strong hybridization with O-$p$ orbitals and the resulting finite interlayer coupling. The projected density of states (PDOS) of Co-($e^\prime_{\rm g}, a_{\rm 1g}, e_{\rm g}$) and O-$p$ orbitals are plotted in Fig.~\ref{fig:Fig1}(c). The PDOS shows a clear splitting between ($e^\prime_{\rm g}, a_{\rm 1g}$) and $e_{\rm g}$ states. In this nonmagnetic solution, all six ($e^\prime_{\rm g}, a_{\rm 1g}$)- and a single $e_{\rm g}$-derived bands are occupied as shown in Fig.~\ref{fig:Fig1}(b) and (c), confirming the charge status of $d^7$ for Co$^{2+}$ ions as previously reported \cite{Henry2006}.
We mention that Co$^{2+}$ ions were experimentally reported to stabilize in ($t_{\rm 2g}^5e_{\rm g}^2$, $S=3/2$) high spin configurations with one hole in the $t_{\rm 2g}$ sector.

\section{Metallic character from DMFT calculations without structural relaxation}

\begin{figure*}
\centering
\includegraphics[width=\textwidth]{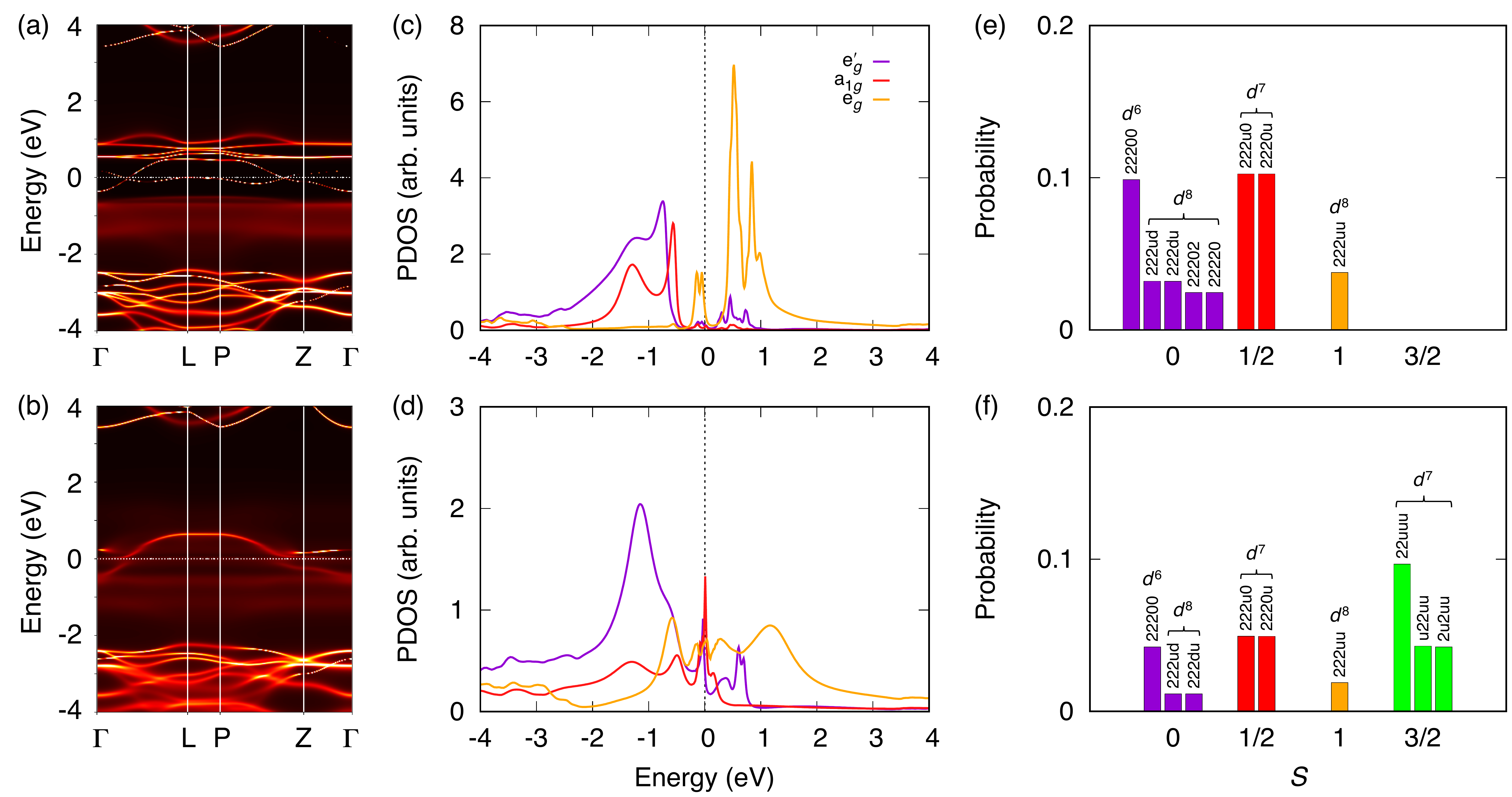}
\caption{(a)-(b) Momentum and frequency dependent spectral function $A({\bf k},\omega)$, (c)-(d) orbital resolved density of states, and (e)-(f)  probabilities of major atomic configurations for ($U, J_{\rm H}$) = (8, 0.8) and (12, 1.2) eV, respectively. The results were obtained in the absence of relaxation of atomic coordinates within DMFT calculations. In the paramagnetic phase, both spin components carry equal and opposite amount of probability. Here, only one spin component is shown. In the atomic configuration 0, $u$, $d$, and 2 stand for an unoccupied state, a singly occupied state with spin up electron, with spin down electron, and doubly occupied state with two opposite spin electrons, respectively.}
\label{fig:Fig2}
\end{figure*}

In order to obtain a more accurate electronic structure of BCAO incorporating the multiplet nature of the local moments, we carried out DMFT calculations. Figure~\ref{fig:Fig2} demonstrates DMFT calculation results with adopting the DFT optimized internal atomic coordinates. Figure~ \ref{fig:Fig2}(a, b) show the momentum and frequency dependent spectral function $A(\bf{k},\omega)$ for two sets of ($U$, $J_{\rm H}$) = (8, 0.8) and (12, 1.2) eV, respectively. A metallic band-like feature is observed in both cases. In the former case, metallic feature remains robust close to the Fermi level, as signified by the sharp band spectra and vanishing imaginary self-energy at $\omega$ = 0 (see App.~\ref{ref:App1} for the imaginary part of the self-energy in the imaginary frequency domain). In the latter case of ($U$, $J_{\rm H}$) = (12, 1.2) eV, the metallicity remains intact, but the enhancement of the correlation-induced scattering rate becomes apparent (check also App.~\ref{ref:App1} for the increase of the imaginary self-energy at $\omega$ = 0).

Figure~\ref{fig:Fig2}(c, d) plots the PDOS of Co $d$-orbitals, where DOS are projected onto trigonally split $e^\prime_{\rm g}$ (violet), a$_{\rm 1g}$ (red), and $e_{\rm g}$ (orange) orbitals. In Fig.~\ref{fig:Fig2}(c), $e_{\rm g}$ states are almost empty except one occupied band, signalling the low-spin configuration and consistent with our nonmagnetic DFT result. On the contrary, in Fig.~\ref{fig:Fig2}(d) $e_{\rm g}$ states become half-filled and $e^\prime_{\rm g}$ noticeably develops a hole, which signifies a low-spin-to-high-spin crossover. Note that in both cases no clear separation between $e^\prime_{\rm g}$ and a$_{\rm 1g}$ states is seen, implying the trigonal distortion within CoO$_6$ octahedra away from the ideal cubic limit is not significant in this geometry. 

The nature of the spin-state crossover can be better understood from the probabilities of atomic configurations obtained from the quantum Monte Carlo impurity solver, as shown in Fig.~\ref{fig:Fig2}(e, f) for ($U$, $J_{\rm H}$) = (8, 0.8) and (12, 1.2) eV, respectively. Note that it suffices to show probabilities for one spin component, thanks to the time-reversal symmetry in the paramagnetic phase. In the former case of $(U, J_{\rm H})$ = (8, 0.8) eV (Fig.~\ref{fig:Fig2}(e)), low-intermediate spin configurations $S=0$, $1/2$, and $1$ prevail, and significant portion of charge fluctuation ({\it i.e.} non-negligible probabilities of $d^{6}$ and $d^8$ configurations) is also noticed. As ($U$, $J_{\rm H}$) is increased to (12, 1.2) eV (Fig.~\ref{fig:Fig2}(f)), suppression of charge fluctuation and occurrence of high-spin $S=3/2$ components can be noticed in comparison to (8, 0.8) eV case. It can be understood that enhanced $U$ reduces charge fluctuation and promotes $d^7$ configurations, while increased high-spin $S=3/2$ probabilities can be attributed to the larger $J_{\rm H}$, both of which contributing to the disturbed metallic character. 

In Fig.~\ref{fig:Fig2}(f), one of the three probabilities of the $d^7$, $S=3/2$ states is larger than the other two (bars in green). The first state with higher probability is identified to have one hole in the nondegenerate $a_{1g}$ orbital, while other two states with smaller and almost same probability are found to have one hole in the doubly-degenerate $e^\prime_{\rm g}$. This is because of slightly higher crystal field level of $a_{\rm 1g}$ orbital compared to the $e^\prime_{\rm g}$ doublet, so that the hole in the $t_{\rm 2g}$ orbital tends to be more located at $a_{\rm 1g}$. We emphasize that while the difference in probabilities is consistent with the nonvanishing (but small) CoO$_6$ trigonal distortion, the finite probabilities of the $S=3/2$ states with $e^\prime_{\rm g}$-hole implies the fact that BCAO may possess an active $L_{\rm eff}=1$ orbital degree of freedom.

\section{Expansion of Co octahedra via DMFT structural relaxation}

\begin{figure}
\centering
\includegraphics[width=0.45\textwidth]{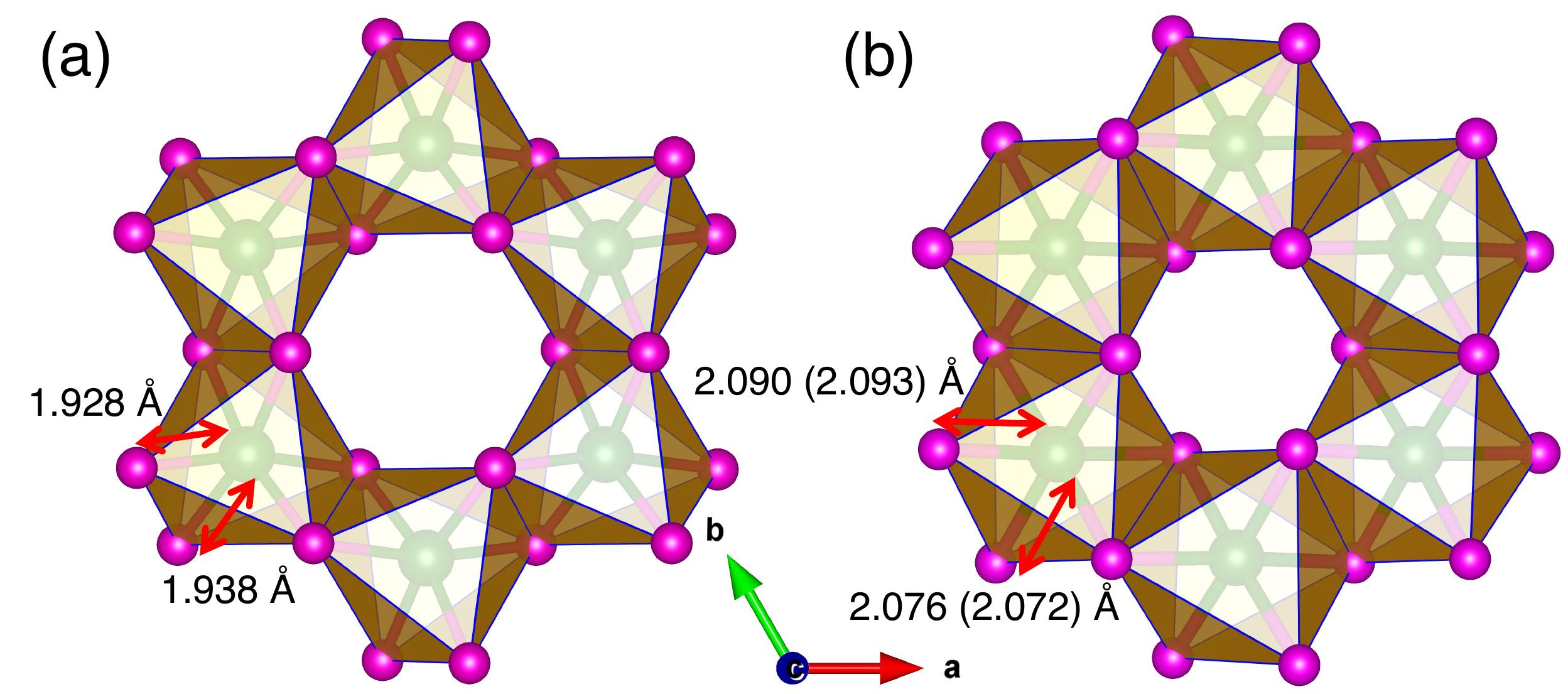}
\caption{(a)-(b) Top view of the single honeycomb layer displays trigonally distorted CoO$_6$ octahedra for optimized crystal structures, obtained using PBEsol functional in DFT and LDA functional in DMFT, respectively. In the right figure, bond lengths are indicated for two sets of ($U, J_{\rm H}$) = (8 (12), 0.8 (1.2)) parameters. For clarity, other honeycomb layers and atoms are removed from the crystal structures.}
\label{fig:Fig3}
\end{figure}

In the previous section, DMFT calculations without relaxation of DFT-optimized internal coordinates showed no signature of Mott insulating phase in BCAO even with higher value of ($U,J_{\rm H}$) parameters, although a couple of signatures of correlations were noticed ({\it e.g.} formation of coherent peak at $E_{\rm F}$ shown in Fig.~\ref{fig:Fig2}(d) and an increase in scattering rate at zero frequency as shown in Fig.~\ref{fig:Fig6}(b) at App.~\ref{ref:App1}). On the contrary, experimentally BCAO has been reported to be a Mott insulator with high-spin Co$^{2+}$ ions. These discrepancies arise due to an overestimated hybridization between the Co $d$ and O $p$ orbitals, which originates from short Co-O bond lengths from DFT optimized structure. This necessitates further structural relaxation within DMFT calculations.

Invoking structural relaxation, internal coordinates of BCAO were optimized within DMFT. Figure~\ref{fig:Fig3} summarizes structural changes caused by DMFT optimizations, where top views of the single CoO$_3$ honeycomb layers from DFT- and DMFT-optimized structures, are presented in panels (a) and (b), respectively. 
By comparing Fig. \ref{fig:Fig3}(a) and (b), we notice that Co-O bond lengths within CoO$_6$ octahedra significantly enhances in the DMFT-relaxed structure compared to the DFT-relaxed one. It is also noticed in Fig.~\ref{fig:Fig3}(b) that small, but non-negligible differences in Co-O bond lengths exist between two DMFT calculations with different choices of Coulomb parameters ($U, J_{\rm H}$) (check the caption of Fig.~\ref{fig:Fig3}). This enhancement of Co-O bond length and the resulting CoO$_6$ octahedral volume expansion is a hallmark of low-to-high spin crossover as reported in many other transition metal oxide compounds \cite{Pickett2008,Chow2008}, which can be attributed to the formation of Co $S=3/2$ high-spin states.

\section{Formation of Mott insulating phase with $S=3/2 \oplus L_{\rm eff}=1$ orbital degree of freedom after DMFT structural relaxation}

\begin{figure*}
\centering
\includegraphics[width=\textwidth]{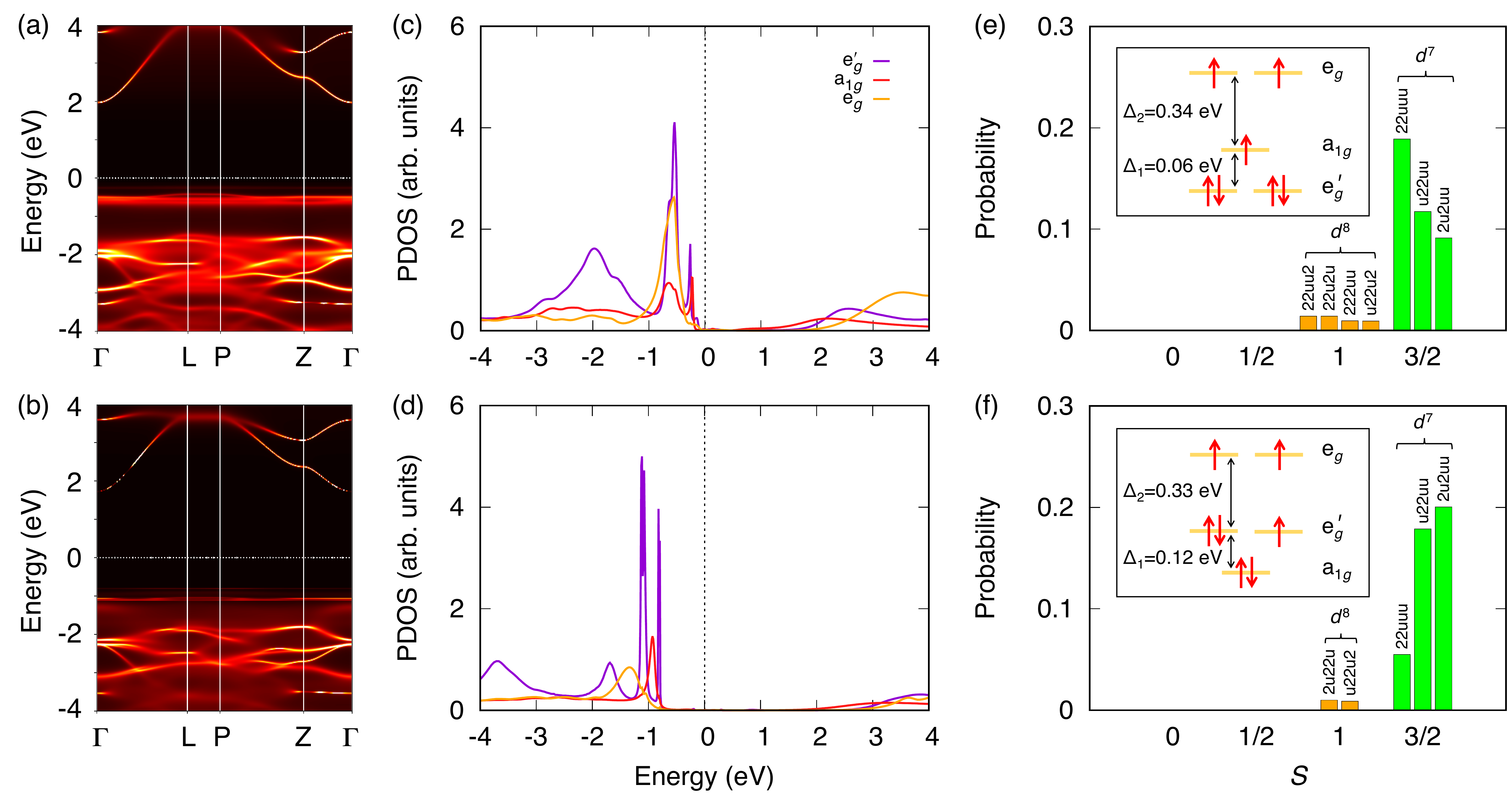}
\caption{Top and bottom panels display (a)-(b) spectral function, (c)-(d) projected density of states, and (e)-(f) atomic configurations with high probability for ($U, J_{\rm H}$) = (8, 0.8) and (12, 1.2) eV, respectively. The insets in (e) and (f) show relative strength of trigonal crystal field splitting of Co$^{2+}$ ions and distribution of seven electrons at $e^\prime_{\rm g}$, $a_{\rm 1g}$, and $e_{\rm g}$ levels. The plots were obtained after DMFT structural relaxations. For details of alphanumeric characters in the atomic configurations, refer to Fig. \ref{fig:Fig2}.}
\label{fig:Fig4}
\end{figure*}

In this section, we present the results with DMFT relaxations of internal coordinates, demonstrating clear Mott-insulating and high-spin characters with active $L_{\rm eff}=1$ orbital degree of freedom. Figure~\ref{fig:Fig4} summarizes the results, where the top and bottom panels present spectral functions for ($U, J_{\rm H}$) = (8, 0.8) and (12, 1.2) eV, respectively. Both the spectral functions, as shown in Fig.~\ref{fig:Fig4}(a, b), exhibit a clear Mott gap and lower Hubbard bands around $-0.5$ eV with respect to the Fermi level.
Dispersive bands are seen in the unoccupied sector, where the onset is about 2 eV above the Fermi level, whose character mostly originates from Ba-$s$ orbitals. 
Increasing the Coulomb parameters ($U, J_{\rm H}$) from (8, 0.8) to (12, 1.2) eV does not change the size of the Mott gap significantly, as shown in Fig.~\ref{fig:Fig4}(a, b), because the unoccupied Ba bands are less affected by the Co $U$ parameter. On the other hand, the gap between the Co-originated lower and upper Hubbard bands is enhanced, as shown in Fig.~\ref{fig:Fig4}(c, d), where the PDOS for $e^\prime_{\rm g}$, $a_{\rm 1g}$, and $e_{\rm g}$ orbitals are shown. It is noticeable that as ($U,J_{\rm H}$) parameters increase, narrow well-localized peaks develop around $-1$ eV.

Figure~\ref{fig:Fig4} (e, f) depict probabilities of major atomic configurations. After DMFT structural relaxations, the Co configuration converges to the $d^7$, $S=3/2$ high-spin states. Now charge fluctuation is suppressed in both Mott-insulating solutions. This is in contrast to the previous DMFT results presented above, where Co-O $d$-$p$ hybridizations were unphysically strong due to underestimated Co-O bond lengths in simple DFT calculations. 

Comparing Fig.~\ref{fig:Fig4}(e, f), we notice an interesting change in the probabilities of the $S=3/2$ states. In Fig. \ref{fig:Fig4}(e), out of three $S=3/2$ states, first one is higher in probability than remaining two. From the analysis of the probability and trigonal crystal fields, we identify the former one as having one hole in the $a_{\rm 1g}$ orbital, while in the other two the hole is located at the $e^{\prime}_{\rm g}$ orbitals. As per the impurity self-energy at infinite frequency, which can be considered as bare crystal fields level, $a_{\rm 1g}$ orbital is higher in energy than $e^\prime_{\rm g}$ states in the case of ($U,J_{\rm H}$) = (8, 0.8) eV. The splitting of the bare energy levels is pictorially depicted in the inset of Fig. \ref{fig:Fig4}(e). Hence, it is energetically favorable to have a hole in the $a_{\rm 1g}$ orbital, resulting in higher probability of the first $S=3/2$ state. As ($U, J_{\rm H}$) parameters increase from (8, 0.8) to (12, 1.2), it shows a different scenario. Now $e^\prime_{\rm g}$ states possess higher probability than $a_{\rm 1g}$ states. Consequently, in the energy level diagram, $a_{\rm 1g}$ lies lower in energy than $e^\prime_{\rm g}$ states (see inset of Fig. \ref{fig:Fig4}(f)), resulting in lower probability of the $S=3/2$ state with the $a_{\rm 1g}$-hole.

Interestingly, enhancing ($U,J_{\rm H}$) does not affect the size of the cubic crystal fields to a great extent, namely the splitting between the $e_{\rm g}$ and the center of mass of the $t_{\rm 2g}$ states; 0.37 and 0.39 eV for ($U, J_{\rm H}$) = (8, 0.8) and (12, 1.2) eV, respectively. Hence, the reversal between $a_{\rm 1g}$ and $e^\prime_{\rm g}$ states in the energy level diagram is only associated with additionally introduced CoO$_6$ trigonal distortion induced by the increase of Coulomb parameters without further change of CoO$_6$ volume, consistent with structural behaviors. As shown in Fig.~\ref{fig:Fig3}(b), an increasing ($U, J_{\rm H}$) parameters cause a slight compression of three Co-O bond lengths by 0.004 \AA, with other three being increased by 0.003 \AA. From close-up views of the CoO$_6$ octahedra between ($U, J_{\rm H}$) = (8, 0.8) and (12, 1.2) eV, we find that CoO$_6$ is compressed along the crystallographic $c$-axis by an amount of approximately 0.011 \AA. This compression on CoO$_6$ results in level crossing between $a_{\rm 1g}$ and $e'_{\rm g}$ states, as shown in the insets of Fig. \ref{fig:Fig4}(e) and (f), and the resulting change in the multiplet probability distribution.

Finally we point out that in both choices of our Coulomb parameters ($U,J_{\rm H}$) = (8, 0.8) and (12, 1.2) eV, all three states remain populated to be the most significant atomic configurations. This is because crystal fields splitting within the Co $t_{\rm 2g}$ orbitals is finite, but remains small (see Fig.\ref{fig:Fig4}(c) and (f) and insets within) in our DMFT structural relaxations. Since the $L_{\rm eff}=1$ orbital degree of freedom remains active, incorporation of atomic spin-orbit coupling at Co sites should entangle the $S=3/2$ and $L_{\rm eff}=1$ degrees of freedom, so that the $J_{\rm eff} = 1/2$, $3/2$, and $5/2$ local moments should emerge as predicted in recent theoretical works \cite{Khaliullin2018,Motome2018,Khaliullin2020,Lui2021}.

\section{Discussion and Summary}

From the analysis of DMFT results, we noticed that Co$^{2+}$ ions form a high spin configuration. The probability analysis clearly revealed the formation of $S=3/2$. In addition, appreciable magnitude of probability of $a_{\rm 1g}$ and $e^\prime_{\rm g}$ confirm that orbital angular momentum is not fully quenched by trigonal crystal fields. Hence, BCAO possesses an active $L_{\rm eff}=1$ orbital degree of freedom and therefore, with incorporation of spin-orbit coupling, Co$^{2+}$ will create the doublet $J_{\rm eff}=1/2$ state as the ground state atomic multiplet. 

Finally, we comment on two things; first, directly including atomic spin-orbit coupling in DMFT calculations and reaching good convergent results could not be achieved, but is currently under study. However, the presence of non-quenched $L_{\rm eff}=1$ orbital degree of freedom in the spin-orbit-coupling-free calculations strongly signals the presence of spin-orbit entanglements when the spin-orbit-coupling is included. Second, our choice of Ising-type density-density Coulomb interactions breaks the full rotational symmetry within the $d$-orbitals that should lead to the lowering of the orbital symmetry in general. It is possible that employing more rotationally symmetric form of Coulomb interactions may further push the system into having more ideally symmetric $L_{\rm eff}=1$ case, which we leave for future studies. 

In conclusion, our study strongly suggests the presence of the pseudospin $J_{\rm eff}=1/2$ moments at Co sites, which is one of the key requirements to realize the long-sought Kitaev quantum spin-liquid phase in BCAO \cite{Cava2020,Armitage2022}. Our study sheds light on the nature of the local magnetic moments and magnetic exchange interactions of BCAO. We finally comment that since other Co-based Kitaev compounds like Na$_3$Co$_2$SbO$_6$ and Na$_2$Co$_2$TeO$_6$ share the same local structures around Co with BCAO \cite{Park2021,Park2022,Stock2020}, our study further suggests the presence of the $J_{\rm eff}=1/2$ moments in other Co-based candidate compounds for Kitaev magnetism. Further theoretical studies on the nature of exchange interactions in above systems may be necessary.

\begin{acknowledgments}
S.S. and H.-S.K acknowledge support from the Korea Research Fellow (KRF) Program and the Basic Science Research Program through the National Research Foundation of Korea funded by the Ministry of Education [Grant No. NRF-2019H1D3A1A01102984 and NRF-2020R1C1C1005900], and also the support of computational resources including technical assistance from the National Supercomputing Center of Korea [Grant No. KSC-2021-CRE-0222]. S.N.-P. was supported by Thailand Science Research and Innovation Fund [Grant No. FRB650082/0227-WU08]. H.-S.K thanks Ara Go, Hosub Jin, Choong H. Kim, Eun-Gook Moon, and Kristjan Haule for fruitful discussions, and  additionally appreciate Asia Pacific Center for Theoretical Physics (APCTP) for its hospitality during completion of this work.
\end{acknowledgments}



\appendix

\section{Details on DMFT results: convergence and imaginary frequency self-energies}
\label{ref:App1}

\begin{figure}
\centering
\includegraphics[angle=0,origin=c,scale=0.35]{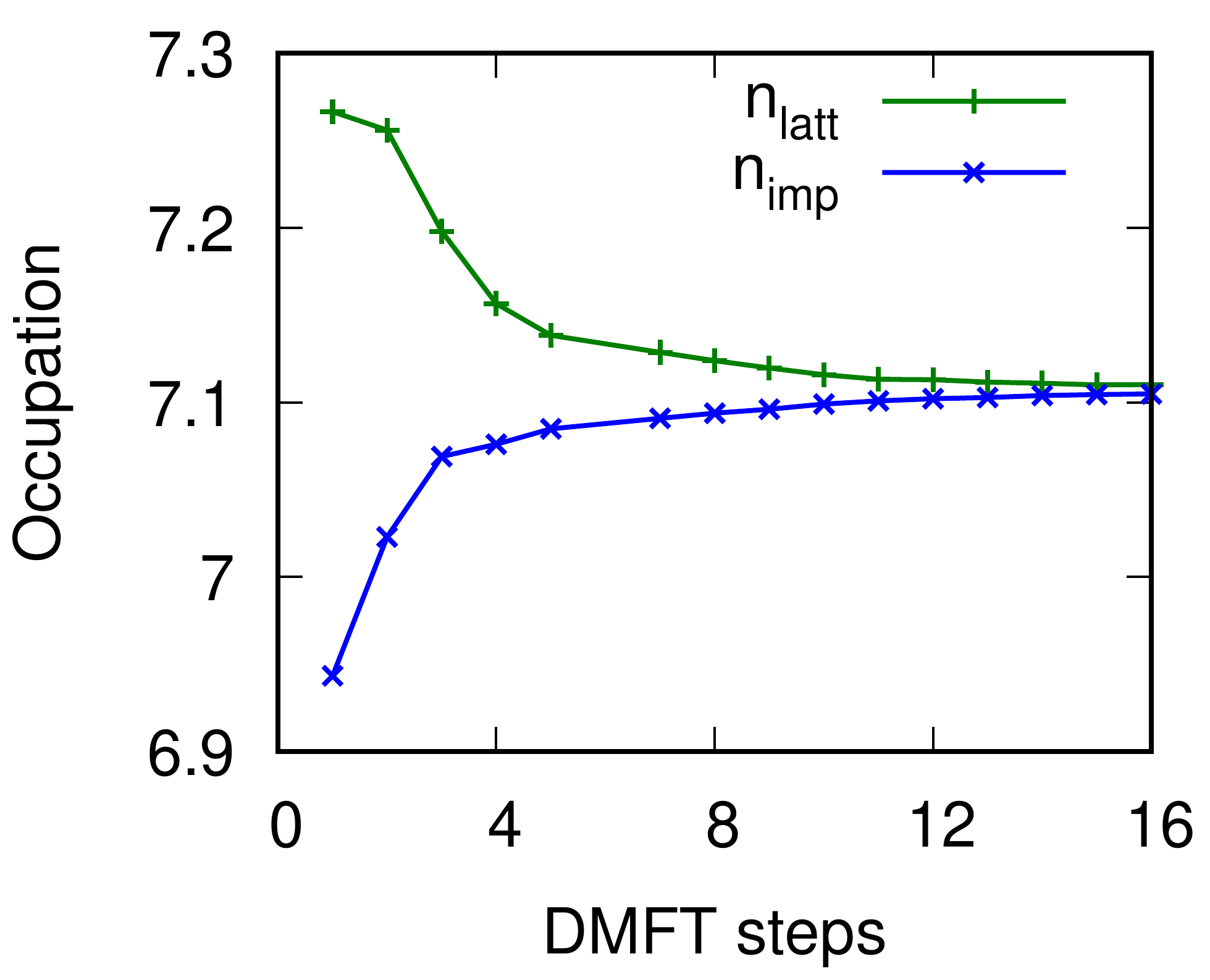}
\caption{Occupation in lattice and impurity for last few DMFT steps with relaxation for $(U, J_{\rm H})$ = (8, 0.8) eV. The occupation is converged to 7.10.}
\label{fig:Fig5}
\end{figure}

In the main text, we have discussed the DMFT results for two sets of $U$ and $J_{\rm H}$ parameters with and without atomic relaxation. To check the convergence of our calculations, we plotted the lattice and impurity occupations as a function of the number of DMFT iterations at ($U$, $J_{\rm H}$) = (8, 0.8) eV with atomic relaxation turned on. Figure~\ref{fig:Fig5} shows the result and we found that the Co-site occupations computed in lattice and impurity problems converge to a value of approximately 7.10, close to the nominal charge of Co$^{2+}$, within 20 full charge-self-consistent iterations. Other choices of Coulomb parameters and structural relaxations showed the same convergence of the lattice and impurity Co charges within 30 full charge-self-consistent iterations at most (not shown). 

\begin{figure}
\centering
\includegraphics[angle=0,origin=c,scale=0.27]{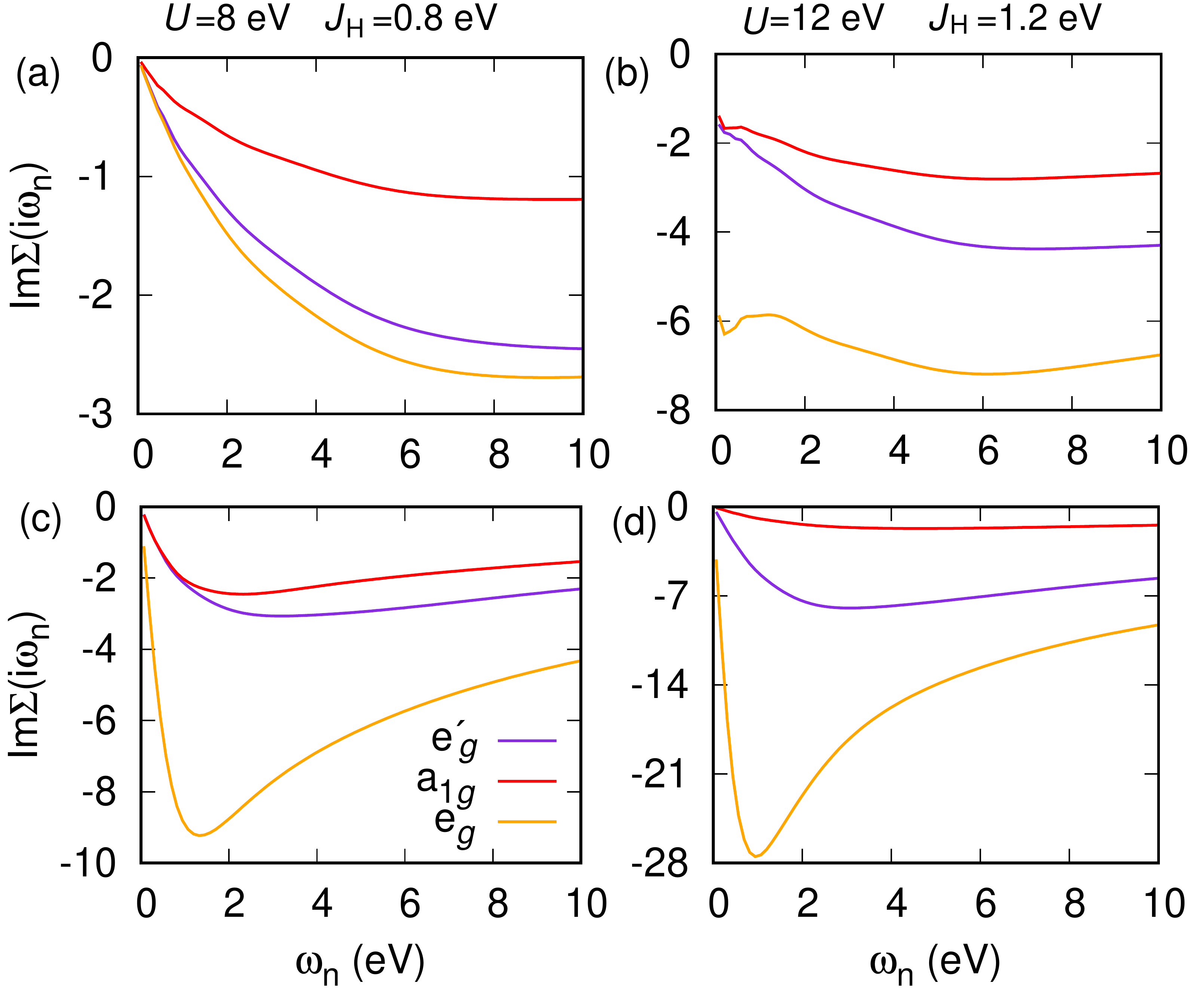}
\caption{Imaginary part of Co self-energy in Matsubara frequency axis for (a)-(b) without DMFT relaxation and (c)-(d) with DMFT relaxation.}
\label{fig:Fig6}
\end{figure}

In Fig.~\ref{fig:Fig6} we present imaginary parts of Co self-energies ($e^\prime_{\rm g}$, $a_{\rm 1g}$, and $e_{\rm g}$ orbitals) in the Matsubara frequency domain. All the results show smooth curves, implying statistical noises within the Monte Carlo samplings were well-cancelled out. Comparing Fig.~\ref{fig:Fig6}(a) and (b), where no further structural relaxation were made in the DFT-optimized crystal structure, it can be seen that electron-electron scattering in the low-energy regime ({\it i.e.} zero-frequency value of the self-energies) becomes enhanced as the Coulomb parameters were increased from (8, 0.8) to (12, 1.2) eV. However, the Mott-insulating state is not captured because there is no sign of pole in the self-energy. Analytically continued real-frequency self-energies also show no pole in their imaginary parts (not shown). On the other hand, self-energies after DMFT structural relaxations as shown in Fig.~\ref{fig:Fig6}(c) and (d) demonstrate signatures of Mott-insulating character in the $e_{\rm g}$-component ({\it i.e.} enhancement in magnitude down to 1 eV and sharp upturn), although the pole in the self-energies is not shown because the chemical potential is not exactly set to the center of the band gap. Analytically continued real-frequency self-energies clearly show poles in their imaginary parts in this case (not shown).

\section{DMFT results with rotationally invariant Coulomb interactions}
\label{ref:App2}

\begin{figure}[H]
\centering
\includegraphics[angle=0,origin=c,scale=0.205]{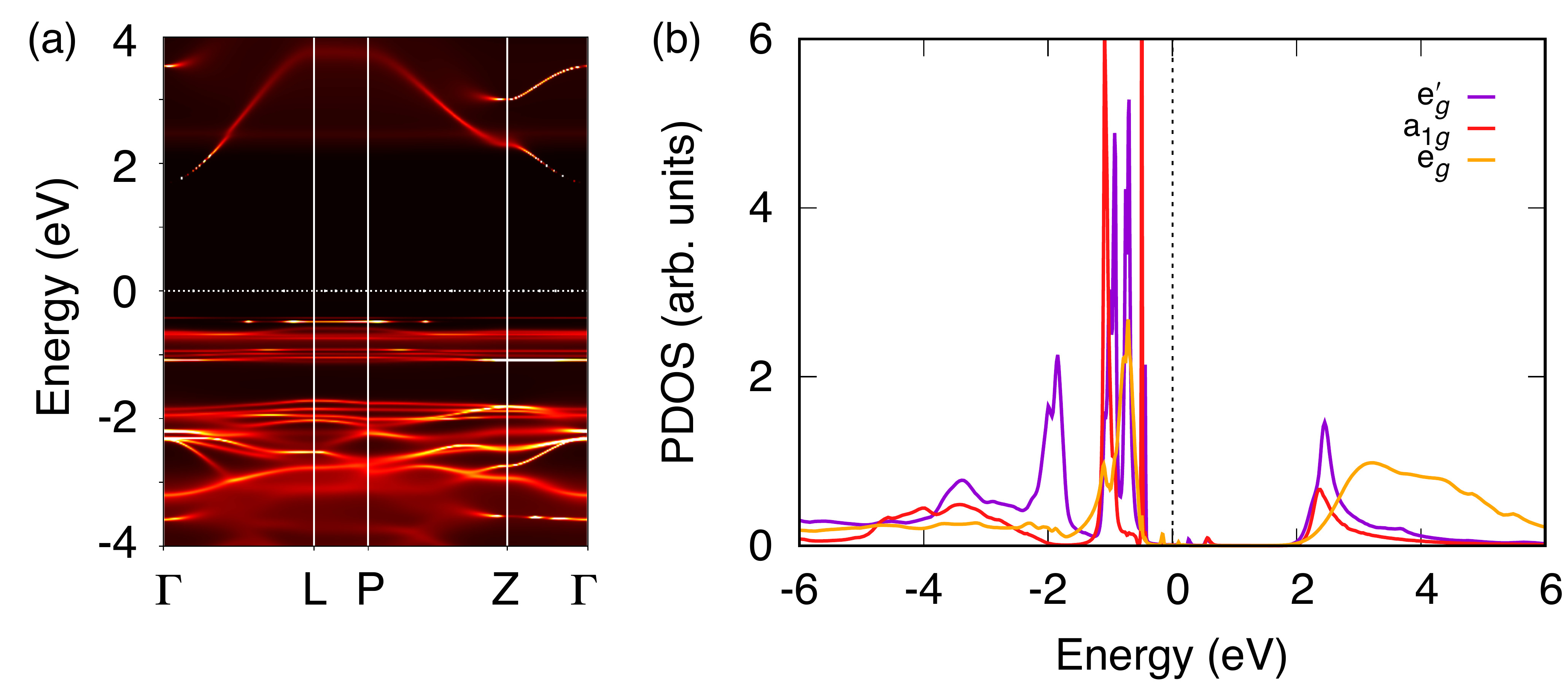}
\caption{(a) Spectral function and (b) orbital resolved density of states of BCAO for $(U, J_{\rm H})=(8, 0.8)$ eV with rotationally invariant Coulomb interaction, respectively.}
\label{fig:Fig7}
\end{figure}

In the main part, we presented all our results using Ising type density-density Coulomb interaction. To check the persistence of $L_{\rm eff}$ = 1 triplet state with the choice of rotationally invariant Coulomb interactions in our DMFT calculations, we computed projected density of states with using full Coulomb interactions including spin exchange and pair hopping terms. Fig.~\ref{fig:Fig7} summarizes the results, where orbital-projected density of states of Co in the presence of full Coulomb interaction is plotted. The PDOS clearly shows $a_{1g}$ and $e^\prime_g$ hole peaks around 2.5 eV above the Fermi level. Most importantly, size of the $e^\prime_g$ hole is twice that of $a_{1g}$. Because the charge and spin configuration of Co remains to be $d^7$ high-spin, this shows that a single hole is equally distributed over the three $t_{\rm 2g}$ orbitals, signifying the robustness of the $L_{\rm eff}$ = 1 triplet. We note in passing that switching Ising to full Coulomb interaction induces a minor change of crystal field splitting size (less than 0.03 eV reduction of the $a_{\rm 1g}$-$e'_g$ splitting), which does not affect our conclusion.

\section{Choice of nominal occupancy in the double counting scheme}
\label{ref:App3}

\begin{figure}[H]
\centering
\includegraphics[angle=0,origin=c,scale=0.2]{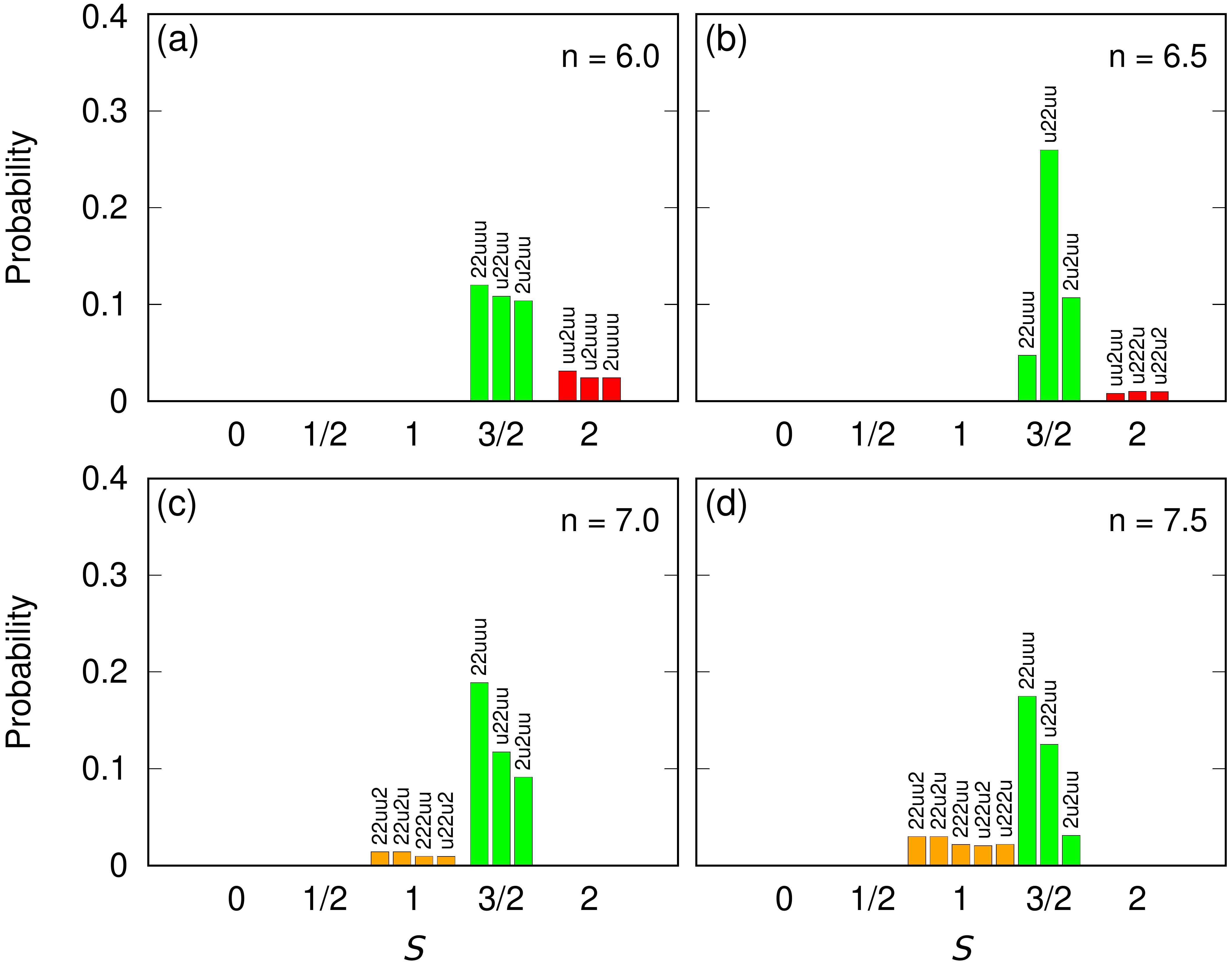}
\caption{(a)-(d) Probability of major atomic configuration of BCAO for $(U, J_{\rm H})=(8, 0.8)$ eV with starting nominal valence $n$ = 6.0, 6.5, 7.0, and 7.5, respectively.}
\label{fig:Fig8}
\end{figure}

All the results presented in our paper were obtained employing the {\it nominal} double-counting (DC) scheme, which is known to be a reasonable choice in many cases. In the {\it nominal} DC scheme, the double counting energy to be subtracted can be expressed in terms of the expected nominal occupancy ($n$) of ionic state of correlated atom in the following way: $V_{DC}=U(n-1/2)-(J/2)(n-1)$ \cite{Haule2010}. Here, $n$ controls the position of correlated bands in the spectrum and can be used as a tuning parameter of the double counting. We started with different choices of nominal occupancy, varying from 6 to 7.5 with the step size of 0.5. The results are shown in Fig. \ref{fig:Fig8}. After 50 full charge self-consistent cycles, lattice and impurity $d$-occupancy in solid converges to 6.79, 6.99, 7.10, and 7.29 for $n$ = 6.0, 6.5, 7.0, and 7.5, respectively. Except for the case of $n$ = 6.0, which gives a metallic solution, the rest three cases result in Mott insulating phases with the $L_{\rm eff}$ = 1 degree of freedom alive ({\it i.e.} the nonzero probabilities of the three $S=3/2$ states). This tells that our conclusion of active $L_{\rm eff}$ = 1 triplet is valid over a range of double counting with 6.5 $<n<$ 7.5. Since the nominal occupancy of Co$^{2+}$ is 7.0, we presented our result with this choice of $n$ in the main text.

\section{Convergence with respect to the choice of k-grid}
\label{ref:App4}

\begin{figure}[H]
\centering
\includegraphics[angle=0,origin=c,scale=0.19]{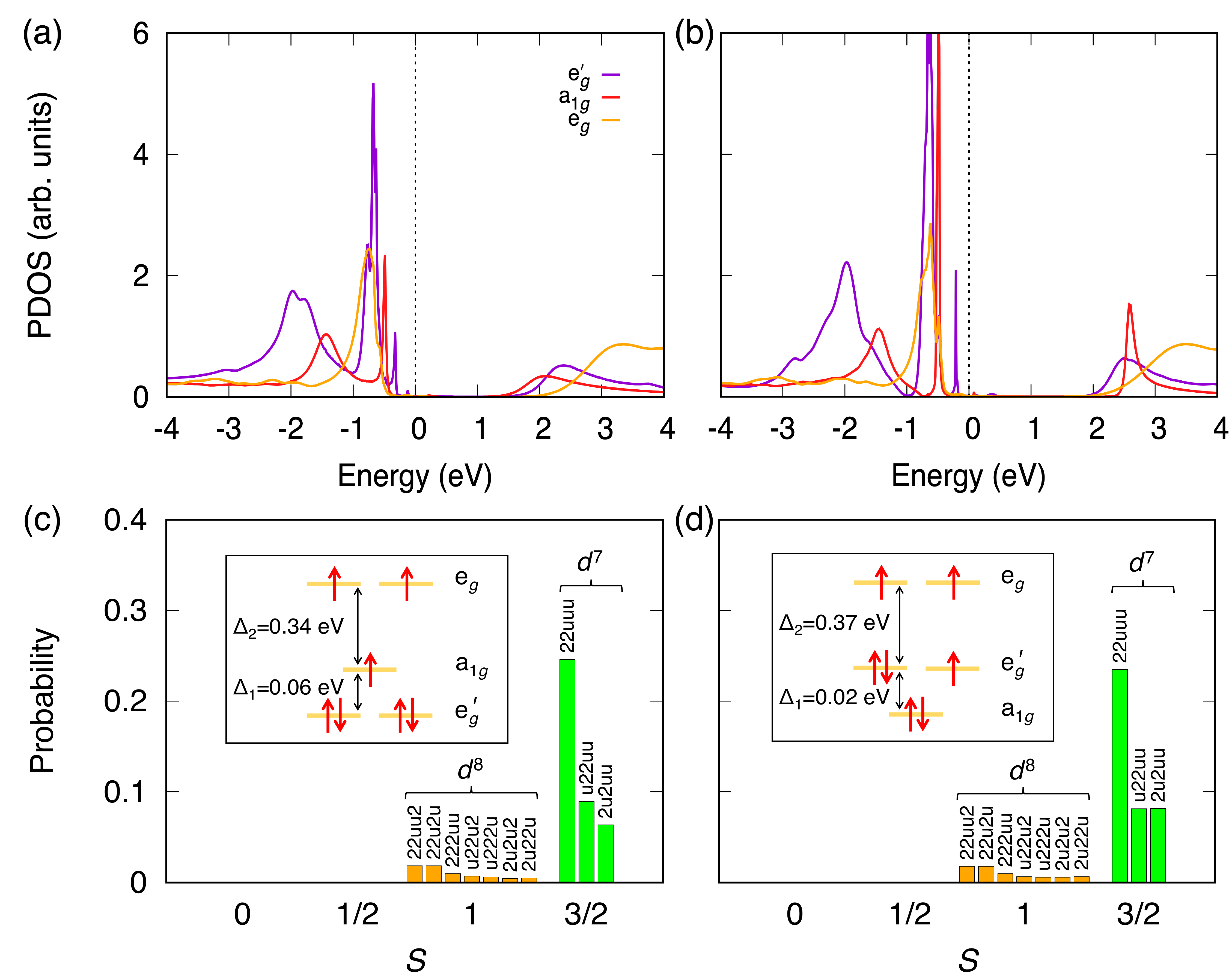}
\caption{(a)-(b) Orbital resolved density of states and (c)-(d) probability of major atomic configurations for $12\times12\times12$, and $14\times14\times14$ $k$-mesh, respectively.}
\label{fig:Fig9}
\end{figure}

The hybridization function may be affected by the density of $k$-grids employed for the numerical integration. Therefore, to check the convergence with respect to the choice of $k$-grids, we have performed additional calculations with employing two denser $k$-meshes, $12\times12\times12$ and $14\times14\times14$ for ($U, J_{\rm H}$)=(8, 0.8) eV at $T$=232K. Figure \ref{fig:Fig9} exhibits projected density of states and probability distributions of atomic multiplets from two different $k$-samplings. We find that, while the probability difference between the [22uuu] and the other two ([u22uu] and [2u2uu]) configurations is somewhat enhanced as we increase the $k$-sampling from 7$^3$ to 12$^3$ (also, see Fig. \ref{fig:Fig4}(e)), but the difference tends to saturate at higher $k$-sampling of 14$^3$. Because the results tend to converge to yield nonzero finite probabilities for all three states within the $L_{\rm eff}$ = 1 triplet as we increase the $k$-sampling, we conclude that $L_{\rm eff}$ = 1 degree of freedom is likely to survive in this system.

We note in passing that, the crystal field levels for $a_{1g}$ and $e_{g}^\prime$ are switched as the $k$-grid is enhanced from 12$^3$ to 14$^3$, so that $a_{1g}$ becomes lower in energy than $e_{g}^\prime$ states at 14$^3$ $k$-grid. This behavior needs further investigations with more accurate computational settings and resources, but implies that the splitting between $a_{1g}$ and $e_{g}^\prime$ is small, therefore is likely to be overwhelmed by the presence of spin-orbit coupling to yield the spin-orbit-entangled $J_{\rm eff}=1/2$ moments.

\section{DMFT calculation results at low temperature}
\label{ref:App5}

\begin{center}
\begin{figure}[H]
\includegraphics[angle=0,origin=c,scale=0.27]{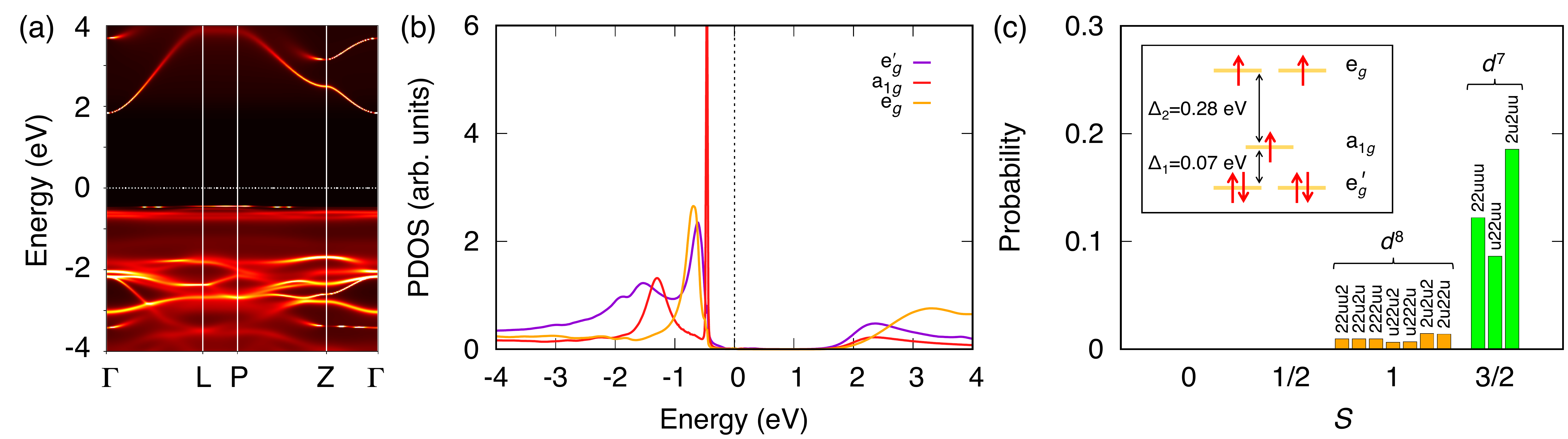}
\caption{(a)-(c) DMFT spectral function, projected density of states, and probability of major atomic configurations, respectively at T=50 K.}
\label{fig:Fig10}
\end{figure}
\end{center}

In order to see whether our DMFT calculation results and conclusions still remain valid at low temperature, we performed calculations at T=50 K. The results are shown in Fig. \ref{fig:Fig10} for ($U,J_{\rm H}$)=(8, 0.8) eV with Ising type Coulomb interaction. We find that BCAO still retains Mott insulating character at low temperature (see left and middle panels). There is some quantitative difference between the results from $T$ = 50 and 232 K (for example differences in peak shape below the Fermi level, which was likely caused in analytic continuations of self-energies), but most importantly, the presence of $a_{\rm 1g}$ and $e'_g$ holes can also be seen at $T$ = 50 K. The probability distribution of multiplet states in the rightmost panel also show that the presence of the $L_{\rm eff}$ = 1 triplet remains robust in the lower temperature result. Crystal field levels, depicted in the inset of the rightmost panel, also show minor change as the temperature is lowered. Since PDOS and the crystal field level did not show significant changes as $T$ is lowered from 232 to 50K, we believe that orbital polarization within the Co $t_{\rm 2g}$ and the resulting reduction of the $L_{\rm eff}$ = 1 triplet would not occur even in the zero-temperature limit.

\onecolumngrid
\twocolumngrid

\bibliography{bcao}

\end{document}